\documentclass[prl,aps,superscriptaddress,twocolumn,longbibliography]{revtex4-1}
\usepackage{graphicx} 
\usepackage{amsmath}
\usepackage{amssymb}
\usepackage{amsfonts}
\usepackage{bbold}
\usepackage{bm}
\usepackage{color}
\usepackage{braket}
\usepackage[colorlinks,linkcolor=blue,citecolor=blue,allcolors=blue]{hyperref}

\begin{document}

\title{Nonlinear planar magnetotransport as a probe of the topology of surface states}

\author{Maria Teresa Mercaldo}
\affiliation{Dipartimento di Fisica ``E. R. Caianiello", Universit\`a di Salerno, IT-84084 Fisciano (SA), Italy}

\author{Mario Cuoco}
\affiliation{CNR-SPIN, I-84084 Fisciano (Salerno), Italy, c/o Universit\'a di Salerno, I-84084 Fisciano (Salerno), Italy}

\author{Camine Ortix}
\email{cortix@unisa.it}
\affiliation{Dipartimento di Fisica ``E. R. Caianiello", Universit\`a di Salerno, IT-84084 Fisciano (SA), Italy}

\date \today


\begin{abstract}
It has been recently established that transport measurements in the nonlinear regime can give direct access to the quantum metric (QM): the real part of the quantum geometric tensor characterizing the geometry of the electronic wavefunctions in a solid. In topological materials, the QM has been so far revealed in thin films of the topological antiferromagnet MnBi$_2$Te$_4$ where it provides a direct contribution to longitudinal currents quadratic in the driving electric field. Here we show that the Dirac surface states of strong three-dimensional topological insulators have a QM that can be accessed from the nonlinear transport characteristics in the presence of an externally applied planar magnetic field. A previously unknown intrinsic part of the longitudinal magnetoconductivity carries the signature of the QM while coexisting with the extrinsic part 
responsible for the so-called bilinear magnetoelectric resistance. 
We prove that the QM-induced nonlinear magnetotransport carries specific signatures of single Dirac cones. This allows to use it as an efficient diagnostic tool of the bulk topology of three-dimensional non-magnetic insulators.
\end{abstract}

\maketitle

\paragraph{Introduction --} Topological insulators (TIs)  are materials that are insulating in their bulk but allow for 
electrical conduction along their surfaces~\cite{has10,qi11}. This macroscopic property is the consequence of the topology of the ground state of the insulator, and the essence of the bulk-boundary correspondence. In a TI the electrical conduction is due to surface electronic modes that are
``anomalous": the surface of strong three-dimensional TIs~\cite{fu07,moo07,fu07b} feature single Dirac cones violating the fermion doubling theorem~\cite{nie81}. 

The electronic wavefunctions of these anomalous surface states additionally carry non-trivial geometric properties encoded in the quantum geometric tensor~\cite{pro80,mer21,tor23}. Its imaginary part -- the Berry curvature -- 
can be non-vanishing since any centrosymmetry of the bulk, when present, is naturally broken at the surfaces. 
In materials whose surface point group does not contain evenfold rotation symmetries, such as Bi$_2$Se$_3$ and related compounds~\cite{zha09,xia09,che09,hsi09,fu09}, 
alternating positive and negative regions of Berry curvature then exist. These can be potentially probed in anomalous planar Hall experiments~\cite{bat21,cul21,wan23} or, in the absence of external magnetic fields, by second-order nonlinear transport 
including the nonlinear Hall effect with time-reversal symmetry~\cite{sod15,du19,ort21,du21}. 
The latter
has been observed 
in Bi$_2$Se$_3$ thin films~\cite{he21}
and connected to
an higher-order moment of the Berry curvature, dubbed Berry curvature triple, which is
responsible for 
non-linear side-jumps and skew scattering processes~\cite{mak24}.

Recent studies have shown that the real part of the quantum geometric tensor, {\it i.e.} the quantum metric (QM) of the electronic wavefunctions, can also give rise to nonlinear transport phenomena. The QM can trigger, via electric field-induced correction to the Berry curvature~\cite{gao14}, a non-linear anomalous Hall effect predicted  in antiferromagnetic metals~\cite{liu21,wan21} and Eu-based Zintl compounds~\cite{zha24}, 
and an intrinsic nonlinear planar Hall effect proposed in two-dimensional transition metal dichalcogenides~\cite{hua23}.
Furthermore, a QM-induced velocity yields a dissipative intrinsic second-order longitudinal current~\cite{das23,kap24} $j_{a}=\chi_{aaa} E_a^2$, with $E_a$ the driving electric field, that has been observed~\cite{wan23exp,gao23} in 
MnBi$_2$Te$_4$ and at oxide heterostructures~\cite{sal24}.  
The corresponding nonlinear longitudinal conductivity reads~\cite{das23} 
\begin{equation}
\chi_{aaa}^{\text{QM}}=\dfrac{e^3}{\hbar}\sum_n \int \dfrac{d^d {\bf k}}{(2 \pi)^d} f_n({\bf k}) \Lambda_{aaa}^{n}({\bf k}),  
\label{eq:eq1}
\end{equation}
where $f_n$ is the Fermi function of the $n$-th 
band and $\Lambda_{aaa}^{n}({\bf k})=3/2 \partial_{k_a} G^n_{aa}$ is the dipole density of the band-energy normalized quantum metric (BNQM) $G^n_{aa}$  
\footnote{The band-energy nornalized quantum metric can be expressed in terms of the interband Berry connection ${\mathcal A}^{nm}_a=i \langle u_{\bf k}^n | \partial_{k_{a}}  u_{\bf k}^m \rangle$, $| u_{\bf k}^n \rangle$ being the Bloch wave of the $n$-th band,  as $G_{ab}^n= 2 \textrm{Re} \sum_{m \neq n} \dfrac{{\mathcal A}^{nm}_a {\mathcal A}^{mn}_b} {\epsilon_n - \epsilon_m}$ with $\epsilon_i$ the Bloch bands. Note that for a generic two-band system the band-energy normalized quantum metric can be related to the quantum metric $g_{ab}=\mathrm{Re} \left[\braket{\partial_a u_{\bf k}^n | \partial_b u_{\bf k}^n} - \braket{\partial_a u_{\bf k}^n | u_{\bf k}^n} \braket{ u_{\bf k}^n | \partial_b u_{\bf k}^n } \right]$ via the relation $G_{ab}=\pm g_{ab} / (2 h)$ where $h$ is the norm of the Hamiltonian vector $\mathbf{h}$  in ${\mathcal H}={\mathbf h} \cdot {\boldsymbol \sigma} + h_0 \sigma_0$ with ${\boldsymbol \sigma}$ the Pauli matrix vector and $\sigma_0$ the identity matrix.}.  

\begin{figure}[tbp]
    \begin{center}
         \includegraphics[width=0.99\columnwidth]{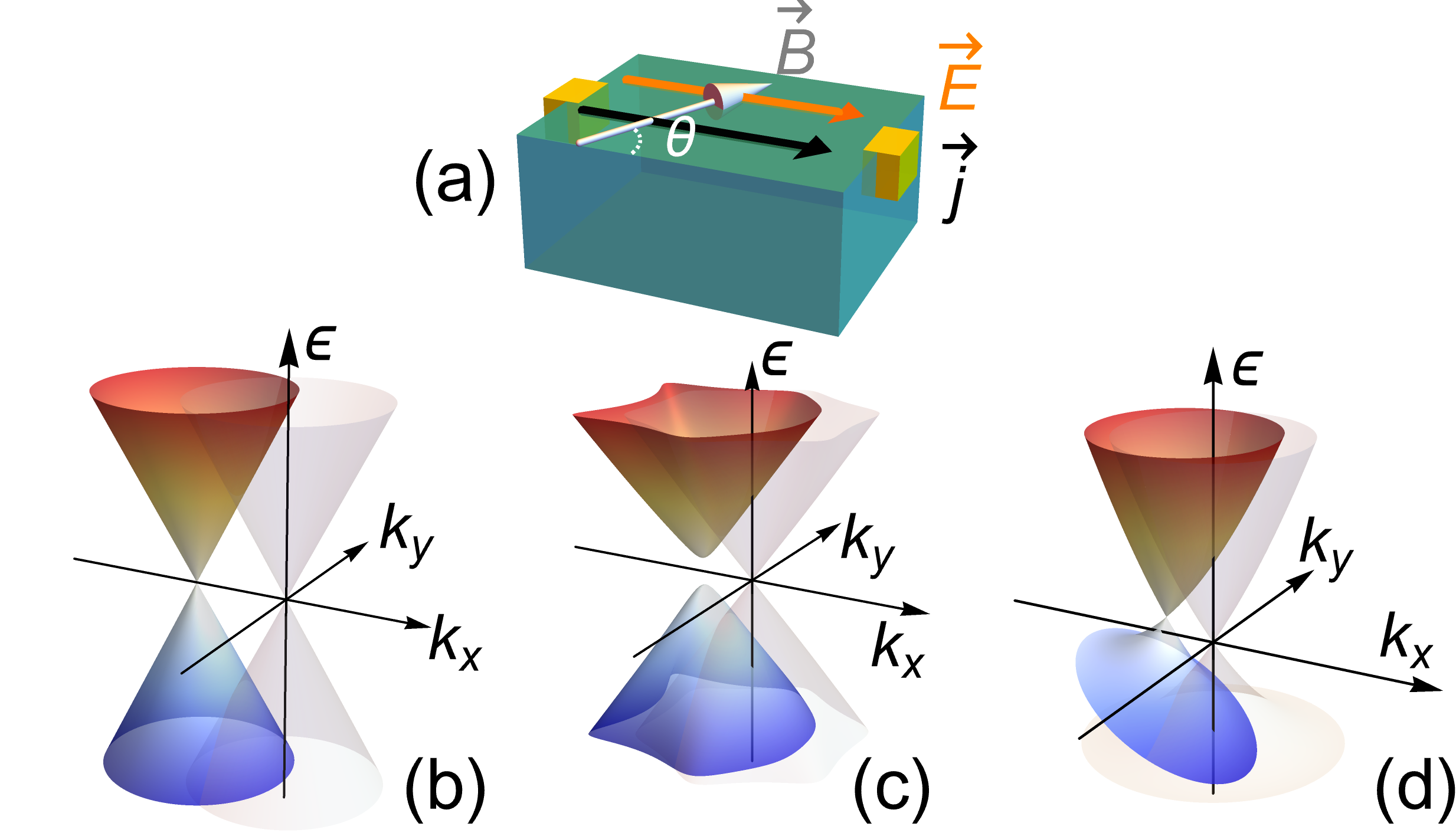}
    \caption{(a) Schematics of the electrical measurement configuration for the QM-induced longitudinal nonlinear currents $j_a=\chi_{aaa} E_a^2$ in a TI subject to a planar magnetic field. A linearly dispersing surface Dirac cone is shifted in momentum space by a planar magnetic field (b). Crystalline anisotropies in trigonal TIs cooperate with the Zeeman coupling to create 
a tilted massive Dirac cone (c). Particle-hole symmetry breaking terms in cubic TIs lead to a massless tilted Dirac cone (d).}
    \label{fig:fig0}
    \end{center}
\end{figure}

The aim of this Letter is to show that the Dirac electrons at the surface of time-reversal invariant TIs feature in their planar magnetotransport [c.f. Fig.~\ref{fig:fig0}(a)] the intrinsic QM-induced nonlinear conductivity of Eq.~\eqref{eq:eq1}. 
The effect of a planar magnetic field on surface Dirac cones is to simply shift them in momentum space [c.f. Fig.~\ref{fig:fig0}(b)]. However, in conjunction with the crystalline anisotropy effects of trigonal TIs, a planar magnetic field yields a tilted massive Dirac cone [c.f. Fig.~\ref{fig:fig0}(c)] whose QM gives a finite intrinsic nonlinear conductivity. 
We also predict a QM-induced nonlinear conductivity in TIs without surface Berry curvature~\cite{waw22}, for instance $(001)$-grown strained bulk HgTe~\cite{bru11}. In this case particle-hole symmetry breaking leads to tilted massless Dirac cones [c.f. Fig.~\ref{fig:fig0}(d)] and non-trivial QM properties. 
We show that a generic fingerprint of single Dirac cones is the absence of sign changes in the intrinsic contribution to the nonlinear magnetoconductivity, contrary to the case of conventional, topologically trivial, spin-orbit coupled surface states.
Our results thus prove 
the capability of quantum metric-induced nonlinear transport in detecting the bulk ${\mathbb Z}_2$ topology of time-reversal symmetric three-dimensional insulators.

\begin{figure}[tbp]
    \begin{center}
         \includegraphics[width=0.99\columnwidth]{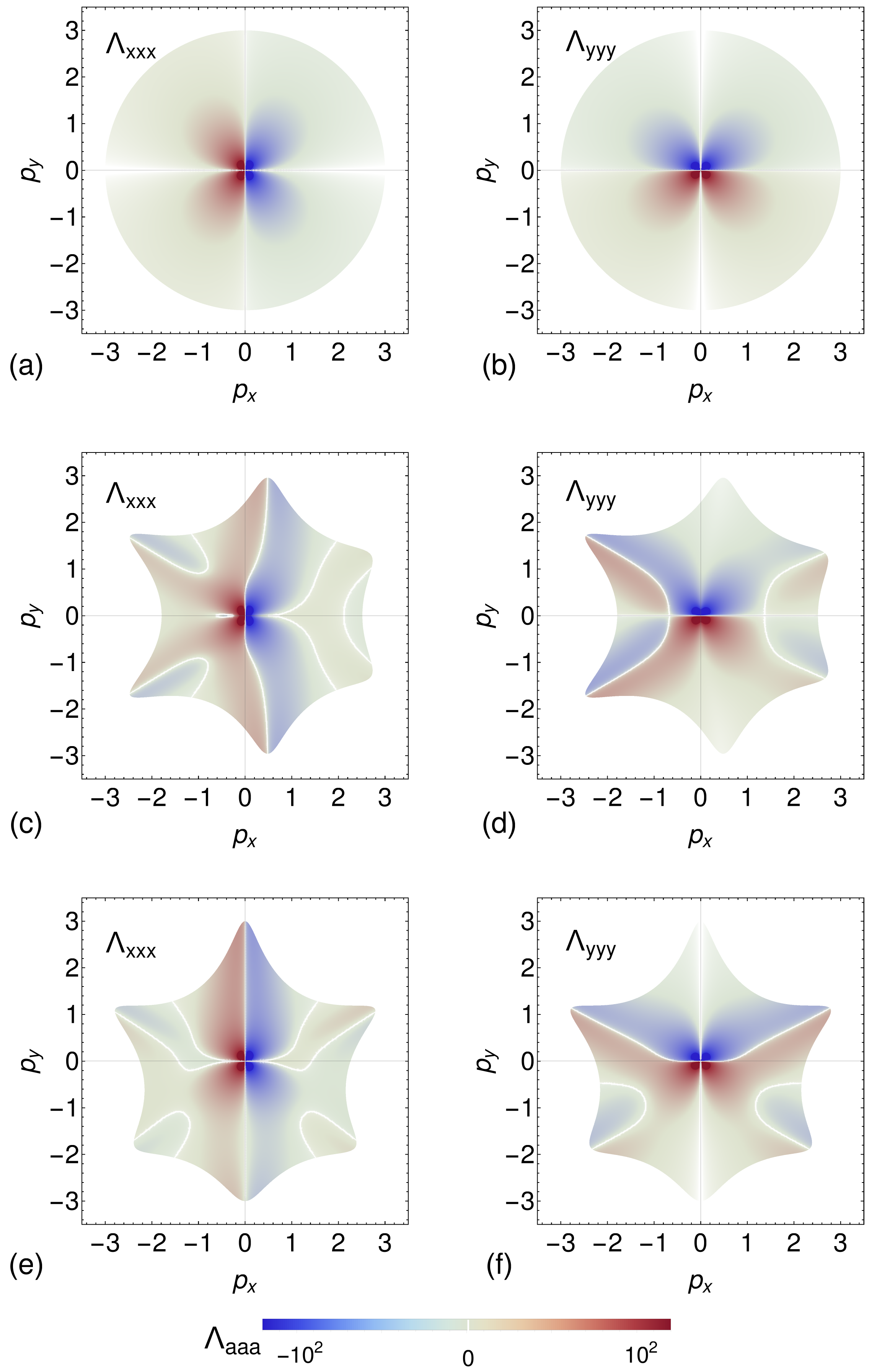}
    \caption{Density plots of the BNQM dipole components $\Lambda_{xxx,yyy}(\mathbf{p})$ {in the plane of the shifted momenta $p_{x,y}$. Momenta are measured in units of a reference momentum $k_0$, which we introduce to write the surface state
Hamiltonian in dimensionless form. The  BNQM dipole is then measured in units of $1/(\hbar v_F k_0^4)$.}
    The top panels (a),(b) are for a Dirac cone where crystalline anisotropies are neglected. The middle (c)-(d) and bottom panels (e)-(f) are with hexagonal warping assuming a planar magnetic field in the $\hat{y}$ and $\hat{x}$ direction respectively. We have considered a Fermi energy $\epsilon_F=3~\hbar v_F k_0$, the warping parameter $\lambda=0.2~\hbar v_F / k_0^2$ and the magnetic field amplitude $B=0.5~\hbar v_F k_0$.}
    \label{fig:fig1v1}
    \end{center}
\end{figure}

\paragraph{TIs in the Bi$_2$Se$_3$ materials class -- }
We first demonstrate that the QM generates nonlinear longitudinal currents 
in trigonal TIs, including Bi$_2$Se$_3$, Bi$_2$Te$_3$ and Sb$_2$Te$_3$.
As mentioned above, the effect of a planar magnetic field on a rotational symmetric Dirac cone is 
a simple shift
in momentum space according to ${\mathcal H}_{\text{TI}}= \hbar v_F \left(k_x \sigma_y - k_y \sigma_x \right) +  {\bf B} \cdot {\boldsymbol \sigma} $. Even though the QM, the BNQM and the BNQM dipole densities 
are all finite, the QM-induced nonlinear conductivity $\chi_{aaa}^{\text{QM}}$ is vanishing. 
The BNQM dipole components $\Lambda_{aaa}$, with $a=x,y$, are odd [c.f. Fig.~\ref{fig:fig1v1}(a),(b)] in the shifted momenta  $p_x=k_x + B_y / (\hbar v_F)$ and $p_y=k_y -  B_x / (\hbar v_F)$, precisely as in a time-reversal symmetric system with the Fermi lines that preserve their circular shapes.  
However, the situation drastically changes when crystalline anisotropy terms are accounted for.

The ${\mathcal C}_{3v}$ symmetry at the TI surfaces allows for an hexagonal warping term~\cite{fu09} ${\mathcal H}_{w}=\lambda (k_{+}^3 + k_{-}^3) \sigma_z / 2$  where $k_{\pm}=k_{x} \pm i k_y$ and we considered, without loss of generality, one of the three vertical mirror symmetry to be ${\mathcal M}_x$ sending $x \rightarrow -x$. 
This  leads to an out-of-plane tilt of the spin textures which then form alternating meron and antimeron wedges respecting the trigonal symmetry of the material surface~\cite{les23}. The appearance of these three-dimensional spin textures has a twofold effect. 
First, it equips the topological surface states with a finite Berry curvature leading to quantum frequency doubling~\cite{he21}.
Second it leads to longitudinal nonlinear currents associated with the nonlinear Drude conductivity 
\begin{equation}
\chi_{aaa}^{\text{NLD}}=-\dfrac{e^3 \tau^2}{\hbar^3} \sum_n \int \dfrac{d^d {\bf k}}{(2 \pi)^d} f_n({\bf k}) \dfrac{\partial^3 \epsilon_n({\bf k})}{\partial k_{a}^3}, 
\label{eq:eq2}
\end{equation}
with $\epsilon_n$ the $n$-th Bloch band energy, and $\tau$ the constant relaxation time. In the weak magnetic field regime [see Supplemental Material (SM)], this semiclassical charge current density $j$ grows like $B E^2$, thus defining the experimentally observed bilinear magnetoelectric resistance~\cite{pan18}  
 and a related extrinsic non-linear planar Hall effect~\cite{he19}. 
We now show that the 
hexagonal warping also leads to the non-vanishing of the QM-induced nonlinear conductivity in Eq.~\eqref{eq:eq1}. 

Fig.~\ref{fig:fig1v1}(c),(d) show the momentum space maps of the BNQM dipole density components $\Lambda_{xxx}$ and $\Lambda_{yyy}$ for all occupied states up to a given Fermi level assuming a magnetic field parallel to a vertical mirror plane. In this configuration  the threefold rotation symmetry and the three vertical mirror symmetry of the TI are broken leaving the ${\mathcal M}_x^{\prime}={\mathcal M}_x ~ \Theta$ symmetry (with $\Theta$ time-reversal) as the only residual magnetic point group symmetry. 
Importantly, the BNQM dipole density components  $\Lambda_{xxx,yyy}$ loose their odd parity under momentum exchange. Nevertheless the ${\mathcal M}^{\prime}_x$ symmetry
, which sends $p_y \rightarrow -p_y$, 
implies that  $\Lambda_{xxx}(p_x, p_y)=\Lambda_{xxx}(p_x,-p_y)$ while $\Lambda_{yyy}(p_x,p_y)=-\Lambda_{yyy}(p_x,-p_y)$. Together with the constraint imposed by ${\mathcal M}^{\prime}_x$ on the Zeeman-distorted snowflake Fermi lines, $\epsilon_F(p_x,p_y)=\epsilon_F(p_x,-p_y)$, we have that the BNQM dipole density component  $\Lambda_{yyy}$ averages to zero when summed over all occupied states while $\Lambda_{xxx}$ yields a finite contribution. 

Consider now the case in which the magnetic field is perpendicular to one of the three vertical mirror planes, for instance along the $\hat{x}$ direction. 
The ${\mathcal M}_x$ symmetry guarantees $\Lambda_{xxx}$ to be odd under a $k_x \rightarrow -k_x$ exchange and must sum up to zero over all occupied states, differently from $\Lambda_{yyy}$ that is instead even in $k_x$ [see Fig.~\ref{fig:fig1v1}(e),(f)]. We therefore have that the quantum metric induces non-linear charge currents along the principal crystallographic directions  
only in the presence of a finite transversal magnetic field component.

\begin{figure}[tbp]
    \begin{center}
         \includegraphics[width=0.95\columnwidth]{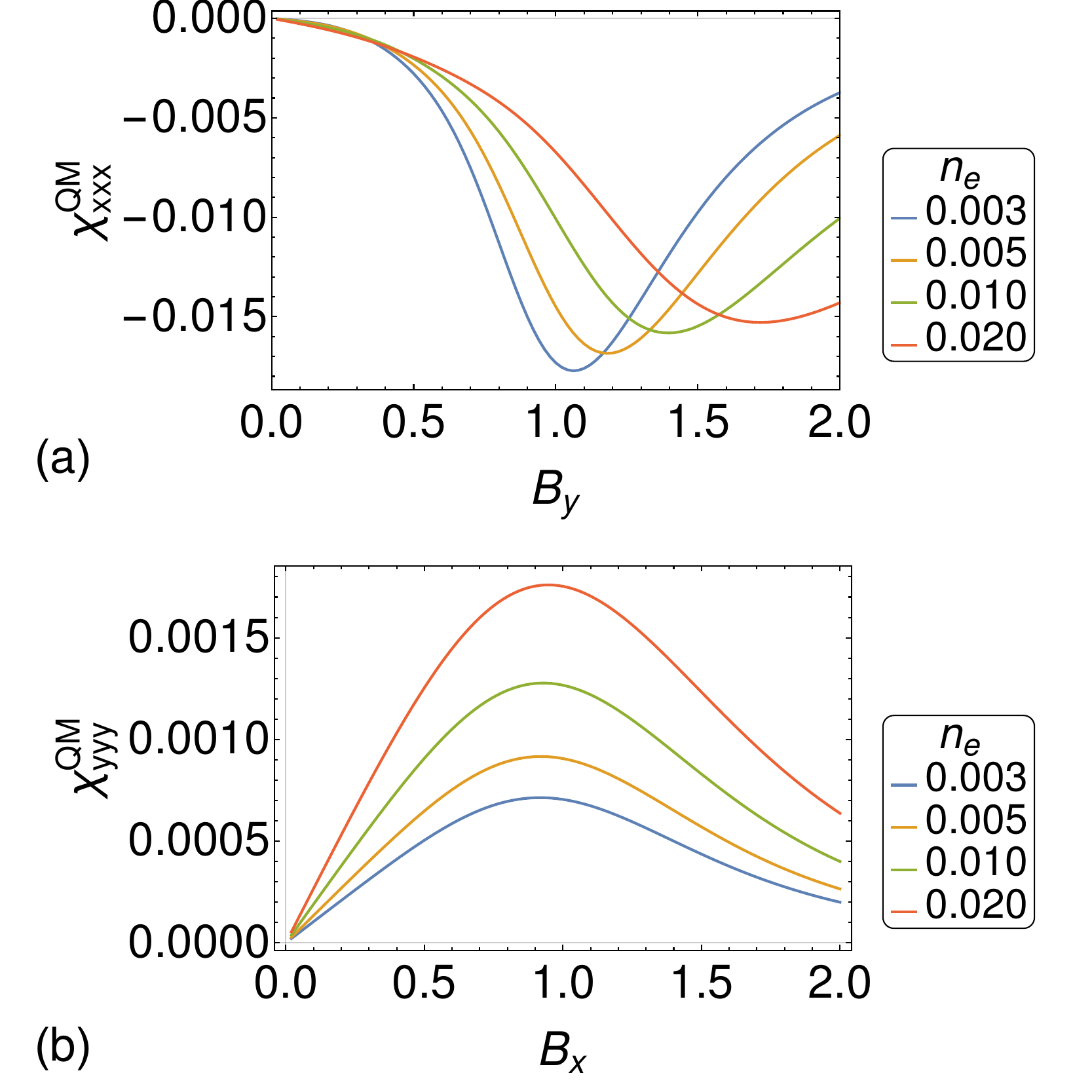}
    \caption{(a) Behavior of the quantum metric induced nonlinear longitudinal conductivity $\chi_{xxx}^\text{QM}$ measured in units of $e^3 / (\hbar^2 v_F k_0^2)$ in a trigonal TI
  as a function the planar  magnetic field (in units of $\hbar v_F k_0$) parallel to one of the vertical mirror planes. (b)  Same for the  nonlinear longitudinal conductivity $\chi_{yyy}^\text{QM}$ with the planar  magnetic field perpendicular to one of the vertical mirror planes. In both panels, the different curves correspond to different fixed values of the surface carrier densities measured in units of $k_0^2$ and the warping parameter is set to $\lambda=0.2~\hbar v_F / k_0^2$. 
  }  
    \label{fig:fig2}
    \end{center}
\end{figure}

\paragraph{QM-induced magnetotransport -- }
Next, we discuss the characteristic features of the nonlinear longitudinal conductivity of Eq.~\eqref{eq:eq1} which is governed by the net BNQM dipole. We first 
consider the nonlinear charge transport response along 
the $\hat{x}$ principal crystallographic direction in the presence of a planar magnetic field ${\bf B} \parallel \hat{y}$. In this geometry, the absence of residual mirror symmetries implies that the topological surface Dirac cones are gapped out with a magnetic field-induced mass $E_g \propto \lambda B^3 / (\hbar v_F)^3$. Furthermore, at very low filling the Fermi lines of the Dirac bands are ellipses centered on the ${\mathcal M}^{\prime}_x$ symmetric line $p_y \equiv 0$ at finite momentum $p_x^0 \propto B^5 \lambda^2 / (\hbar v_F)^7$. This point of the surface Brillouin zone (BZ) differs and does not intersect the lines of zero of the BNQM dipole density separating the nearby high intensity peaks of opposite sign [see SM]. Specifically, $p_x^0$ is at larger and lower momenta for positive and negative values of the magnetic field respectively. 
This offset therefore leads to a net BNQM dipole that switches signs by swapping the direction of the planar magnetic field. 
In the weak field regime, this net BNQM dipole grows, in magnitude, linearly with $B_y$ [see Fig.~\ref{fig:fig2}(a)]. The corresponding nonlinear charge current density $j \propto B E^2$ \cite{Note2},
 which is consistent with the scaling predicted in Ref.~\cite{hua23}.
However, by continuously increasing the magnetic field, we find [c.f. Fig.~\ref{fig:fig2}(a)] for all values of the surface carrier density an optimal strength at which the nonlinear conductivity peaks \cite{Note3}.
This non-monotonous behavior of $\chi_{xxx}^{\text{QM}}$, reminiscent of the behavior of the Berry curvature dipole in strained graphene~\cite{bat19} and oxide heterostructures~\cite{les23,mer23} 
directly derives from the fact that in the $B_y \rightarrow \infty$ limit the quantum metric-induced longitudinal conductivity must vanish. For strong magnetic fields, in fact, the $p_{x,y}$-dependent warping terms coupling to $\sigma_z$ are much smaller than the magnetic field-induced gap. This implies that the surface states are essentially shaped in a rotationally invariant massive Dirac cone largely shifted in momentum space. An explicit calculation of Eq.~\eqref{eq:eq1} proves that the corresponding intrinsic nonlinear conductivity is identically zero. Importantly, this distinguishes
the QM-induced nonlinear magnetoconductivity from the semiclassical Drude contribution since the latter grows monotonically with $B_y$ [see SM]. 
Fig.~\ref{fig:fig2}(b) shows that similar features are found when considering the QM-induced nonlinear magnetoconductivity $\chi_{yyy}^{\text{QM}}$ with a magnetic field along the $\hat{x}$ direction. In this case, the residual vertical mirror symmetry protects the twofold degeneracy at $p_x \equiv p_y \equiv 0$ with the surface Dirac cone that remains gapless at all values of the magnetic field. The vanishing of $\chi_{yyy}^\text{QM}$ in the $B_x \rightarrow \infty$ limit is guaranteed by the fact that in this situation the surface states realize anisotropic Dirac cones with a magnetic field dependent Fermi velocity $v_F^x$, which are not equipped with a net BNQM dipole. 

The QM-induced magnetoconductivity is finite as long as the planar magnetic field is not collinear with the driving electric field. Rotating
the magnetic field in the surface plane, a  $2\pi$ periodic, {\it i.e.} odd under a magnetic field reversal, angular dependence is obtained [see SM]. This is in agreement with the fact that the non-linear conductivity induced by the QM scales with $\tau^0$ [c.f. Eq.~\eqref{eq:eq1}], and must be odd in the planar magnetic field to change sign under a time-reversal operation. 
It is important to note that while the longitudinal nonlinear conductivity $\chi_{xxx}$ is constrained by symmetry to be vanishing in the presence of a collinear magnetic field, the same does not hold for $\chi_{yyy}$. In fact, contributions different from $\chi_{yyy}^{\text{QM}}$ and $\chi_{yyy}^{\text{NLD}}$, for instance due to  side-jump and skew-scattering mechanisms~\cite{du19,ort21,mak24} are allowed in this configuration.

\begin{figure}[tb]
    \begin{center}
         \includegraphics[width=0.99\columnwidth]{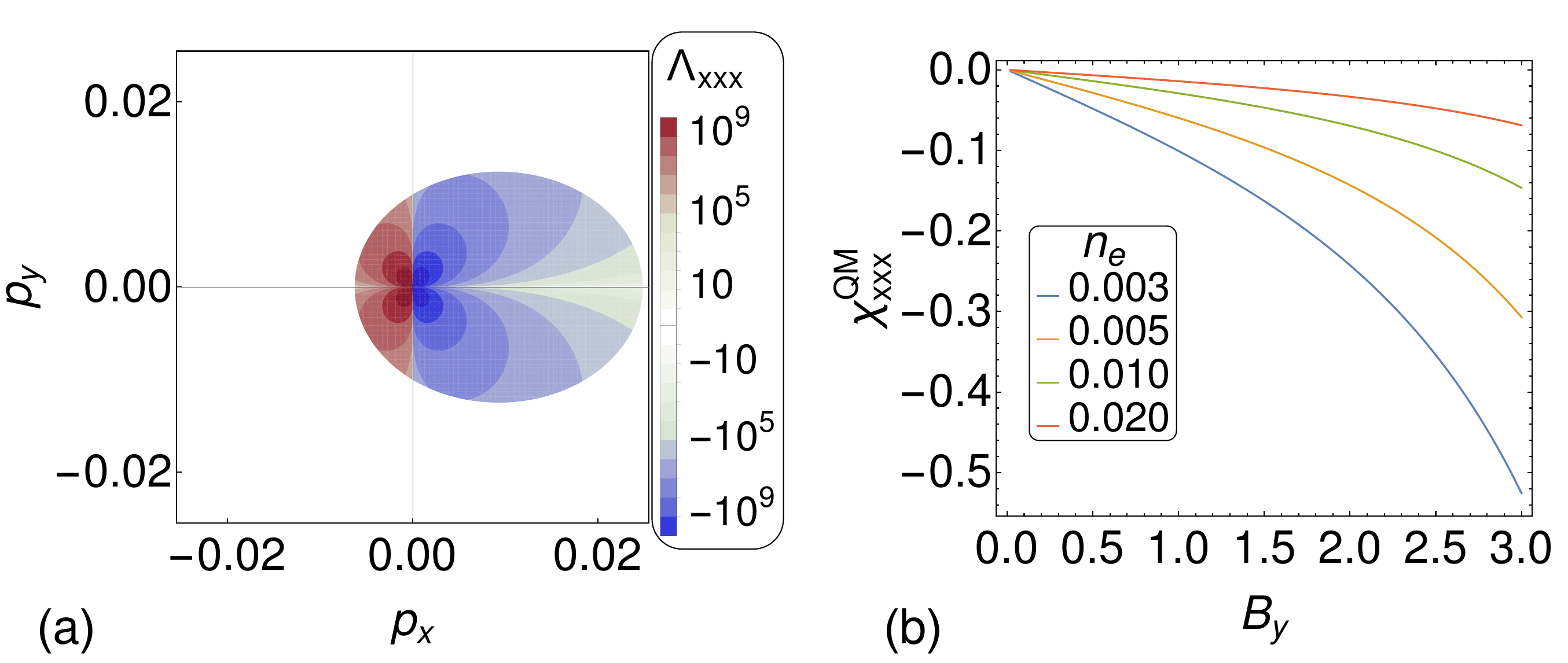}
    \caption{(a) Density plot of the BNQM dipole density component $\Lambda_{xxx}$ measured in units of $1/ (\hbar v_F k_0^4)$ with $k_0$ a reference momentum for a cubic TI grown along the $(001)$ direction in the shifted momenta plane. The planar magnetic field amplitude $B= 3 \hbar v_F k_0$  whereas the parameter measuring the strength of particle-hole symmetry breaking $\alpha=0.1 \hbar v_F / k_0$. We only show the occupied momenta for a Fermi energy $\epsilon_F=0.91 \hbar v_F k_0$. (b) Behavior of the QM-induced nonlinear conductivity measured in units of $e^3 / (\hbar^2 v_F k_0^2)$ as a function of the planar magnetic field strength measured in units of $\hbar v_F k_0$ for different values of the electronic surface density measured in units of $k_0^2$. 
    }
    \label{fig:fig3}
    \end{center}
\end{figure}

\paragraph{HgX compounds --}
Having established the QM-induced nonlinear magnetotransport and its main features in TIs with trigonal crystal symmetry,  we now address the question of whether and how this effect can appear in the series of cubic mercury chalcogenides HgX (X=Te,Se,S) with zincblende crystal structure grown along the $(001)$ direction.
We first recall that HgTe, as well as HgSe, are semimetals which are charge neutral when the Fermi energy is at touching point between the light-hole and the heavy-hole $\Gamma_8$ bands~\cite{dai08,chu11,ort14}. In-plane biaxial strain either splits this fourfold-degenerate Dirac point at the BZ center into a set of four pairs of Weyl nodes sitting at the Fermi level~\cite{rua16}, or lifts the energy degeneracy opening up a topological gap at the Fermi energy~\cite{fu07b,dai08}. 
Metacinnabar -- $\beta$-HgS -- realizes instead a strong 3D TI even in the absence of applied strain~\cite{vir11}. 
To derive the physical properties of the topological surface states in these materials, we use that at the $(001)$ surface the point group is ${\mathcal C}_{2v}$ whose generators are the twofold rotation symmetry along the $z$ axis and a vertical mirror symmetry. A two-dimensional ${\bf k} \cdot {\bf p}$ theory for the $J= \pm 1/2$ surface Kramers doublet 
gives~\cite{ort14}, up to second order in the crystalline momentum, the surface TI Hamiltonian ${\mathcal H}_{\text{HgX}}=\hbar v_F^x k_x \sigma_y - \hbar v_F^y k_y \sigma_x + (\alpha_x k_x^2 + \alpha_y k_y^2) \sigma_0$, with the last two terms that yield particle-hole symmetry breaking. The presence of the two-fold rotation symmetry combined with time-reversal enforces the surface Berry curvature to be locally vanishing~\cite{bat21}. Note also that 
for $v_F^x \equiv v_F^y$ and $\alpha_x \equiv \alpha_y$ the topological surface states display a continuous rotational symmetry. In the remainder we will consider this situation since it does not qualitatively change the nonlinear magnetotransport properties. 

We examine these
assuming a magnetic field parallel to one of the two vertical mirror planes, which we take as ${\mathcal M}_x$ without loss of generality. 
In terms of the shifted momenta $p_{x}=k_{x} + B_{y} / (\hbar v_F)$ and $p_y= k_y$ the effective surface Hamiltonian can be written as  ${\mathcal H}_{\text{HgX}}=\hbar v_F \left(p_x \sigma_y -  p_y \sigma_x \right) - 2 \alpha p_x B_y / (\hbar v_F) \sigma_0 + \alpha (p_x^2 + p_y^2) \sigma_0 + \alpha B_y^2 / (\hbar v_F)^2$. This describes a tilted massless Dirac cone with a Dirac point protected by the residual ${\mathcal M}_y$  vertical mirror symmetry, and a magnetic field-induced tilt.  
The absence of cubic term coupling directly to $\sigma_z$ in the surface Hamiltonian implies that the BNQM dipole component $\Lambda_{xxx}$ possess precisely the behavior displayed in Fig.\ref{fig:fig1v1}(a). However, the magnetic field-induced tilt of the Dirac bands yields a Fermi line that is an ellipse shifted from the origin with its center that lies on the residual mirror invariant line [c.f. Fig.~\ref{fig:fig3}(a)]. This feature directly yields a net BNQM dipole. 
To show this, we write 
Eq.~\eqref{eq:eq1}
as the integral of the BNQM multiplied by the component of the group velocity parallel to the residual mirror plane as $\chi_{xxx}^{\text{QM}}= 3 e^3 / (8 \pi^2)\int d^2 k ~ G_{xx}^{+} ~ v_x ~\delta (\epsilon^{+} - \epsilon_F)$, with $\epsilon^{+}$ the conduction band energy.
For $\alpha B_y / (\hbar v_F)^2 \ll 1$ and at small carrier densities, the Dirac delta can be expanded as 
$\delta (\epsilon^{+} - \epsilon_F)= \delta (\epsilon^{0 +} - \epsilon_F) - 2 \alpha p_x B_y  \partial_{\epsilon^{0 +}} \delta (\epsilon^{0 +} - \epsilon_F) / (\hbar v_F) + \alpha (p_x^2 + p_y^2) \partial_{\epsilon^{0 +}} \delta (\epsilon^{0 +} - \epsilon_F) $ with $\epsilon^{0 +}=\hbar v_F \sqrt{p_x^2+p_y^2}$. Then, we find  the QM-induced nonlinear conductivity 
\begin{equation}
\chi_{xxx}^{\text{QM}} \simeq  - \dfrac{15 e^3 B_y \alpha}{128 \pi \hbar^2 v_F \left(\epsilon_F - \dfrac{\alpha B_y^2}{\hbar^2 v_F^2} \right)^2}.
\label{eq:qmapprox}
\end{equation}
For any magnetic field strength [see Fig.~\ref{fig:fig3}(b)] the intrinsic contribution is enhanced as the surface carrier density is lowered and can reach arbitrary large values. Note that the divergence occurring in the $n_e \rightarrow 0$ limit is regularized by finite temperature effects or
the unavoidable presence of a tiny surface Dirac mass that can originate from hybridization of the Dirac cones at the top and bottom surfaces and even an out-of-plane misalignment of the planar magnetic field [see SM].
Importantly, the Drude semiclassical nonlinear conductivity exhibits a completely different trend:
it has an upper bound at all values of the magnetic field [see SM]. This directly implies that
the nonlinear conductivity will be completely dominated by the QM-induced
intrinsic contribution at low enough electronic densities.


\paragraph{Discussion and conclusions -- } A common trait of the intrinsic QM-induced nonlinear conductivity originating from  single Dirac cones is the absence of a crossover from positive to negative nonlinear magnetoconductivity or {\it vice versa}.
This sharply sets the response of the anomalous surface states of TI apart from the response of conventional surface states appearing in spin-orbit coupled, topologically trivial, materials. Rashba-split surface states display [see SM] 
a sign change of $\chi_{aaa}^\text{QM}$ with an accompanying double peak structure. This is a direct consequence
of the opposite BNQM dipole densities acquired by the inner and outer branches of Rashba surface states.
Put differently, the presence of a sign change in the intrinsic nonlinear longitudinal conductivity directly detects 
 the parity of the surface Kramers pairs at the Fermi level, and hence the bulk ${\mathbb Z}_2$ topology.
We emphasize that the QM-induced nonlinear conductivity always
coexist with the extrinsic nonlinear Drude conductivity of semiclassical origin defining the bilinear magnetoelectric resistance~\cite{pan18}. However, the two different contribution can be parsed using their different scaling with the electronic relaxation time. 
Our results suggest that nonlinear magnetotransport can be used to probe other topological surface states including the Dirac cones violating the fermion multiplication theorem  in topological crystalline insulators~\cite{che19} and hybrid-order topological insulator~\cite{koo20}.

\begin{acknowledgments}
We acknowledge support from the EU Horizon 2020 research and innovation program under Grant Agreement No. 964398 (SUPERGATE), the MAECI project "ULTRAQMAT", and the PNRR MUR project PE0000023-NQSTI (TOPQIN).
\end{acknowledgments}


%

\end{document}


\title{Supplemental Material for: \\
Nonlinear planar magnetotransport as a probe of the topology of surface states} 

\author{Maria Teresa Mercaldo}
\affiliation{Dipartimento di Fisica ``E. R. Caianiello", Universit\`a di Salerno, IT-84084 Fisciano (SA), Italy}

\author{Mario Cuoco}
\affiliation{CNR-SPIN, I-84084 Fisciano (Salerno), Italy, c/o Universit\'a di Salerno, I-84084 Fisciano (Salerno), Italy}

\author{Camine Ortix}
\affiliation{Dipartimento di Fisica ``E. R. Caianiello", Universit\`a di Salerno, IT-84084 Fisciano (SA), Italy}

\maketitle

\section{Additional information on nonlinear magnetotransport in trigonal topological insulators}

\begin{figure}[h]
    \begin{center}
         \includegraphics[width=15.cm]{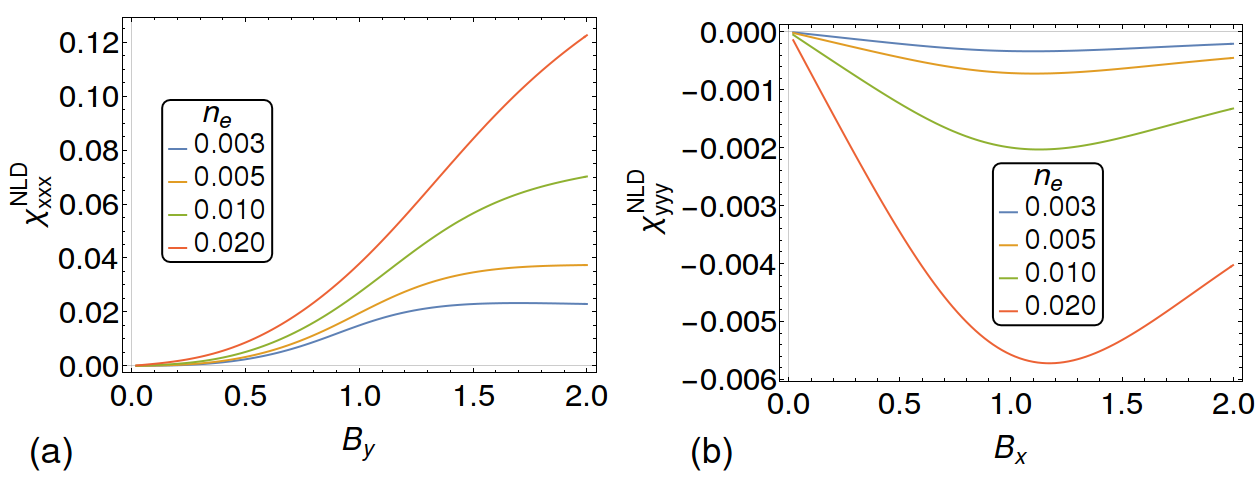}
    \caption{
    Behavior of the extrinsic longitudinal nonlinear Drude conductivity 
    measured in units of $e^3 \tau^2 v_F / \hbar^2$ as a function of the external planar magnetic field strength measured in units of $\hbar v_F k_0$ with $k_0$ a reference momentum. In (a) the magnetic field (given in units of energy) is  parallel and in (b) perpendicular to one of the vertical mirror planes. In both cases the conductivity has been obtained at constant surface carrier  density $n_e$, measured in units of $k_0^2$. 
}
    \label{fig:NLD_BI2SE3}
    \end{center}
\end{figure}

In Fig.\ref{fig:NLD_BI2SE3}  we show the behavior of the extrinsic nonlinear Drude conductivity  $\chi_{aaa}^\text{NLD}$ ($a=x,y$) 
as a function of the applied planar magnetic field, keeping the surface carrier density $n_e$ constant.  For weak magnetic fields the conductivity has a linear behavior, 
and hence the semiclassical current density is $j \propto B E^2$. 
This is the scaling law  that the defines the bilinear magnetoelectric resistance~\cite{pan18}. 
Increasing the applied magnetic field we see deviations from the linear behavior: for the case of the field applied along the $\hat{y}$ direction (i.e. parallel to one of the vertical mirror planes), 
the nonlinear Drude term $\chi_{xxx}^\text{NLD}$ displays a monotonous behavior, with a saturation at high fields, different from the QM-induced intrinsic contribution; when the field is applied along the $\hat x$ direction (i.e. perpendicular to one to the vertical mirror planes), the extrinsic nonlinear Drude conductivity $\chi_{yyy}^\text{NLD}$ instead exhibits a single peak, reminiscent of the behavior of $\chi_{yyy}^{QM}$ discussed in the main part of the manuscript.

The finiteness of the QM-induced intrinsic contribution with a magnetic field parallel to a vertical mirror plane can be understood 
by noticing that in the low density regime, the surface Hamiltonian for topological insulators with trigonal symmetry 
can be approximated as 
\begin{equation}
{\mathcal H}_{Bi{_2}Se{_3}} \simeq \hbar v_F \left(p_x \sigma_y - p_y \sigma_x \right) - \left[\dfrac{\lambda B_y^3}{(\hbar v_F)^3} - 3 \dfrac{B_y^2 p_x \lambda}{(\hbar v_F)^2} \right] \sigma_z 
\end{equation}
From this, one can easily find the gap to be 
$$E_{g} \simeq 2 \sqrt{ \dfrac{\lambda^2 B_y^6}{\left(\hbar v_F \right)^6} \dfrac{1}{1 + 9 \dfrac{\lambda^2 B_y^4}{\left(\hbar v_F \right)^6}}}. $$
\begin{figure}[h]
    \begin{center}
         \includegraphics[width=.75\textwidth]{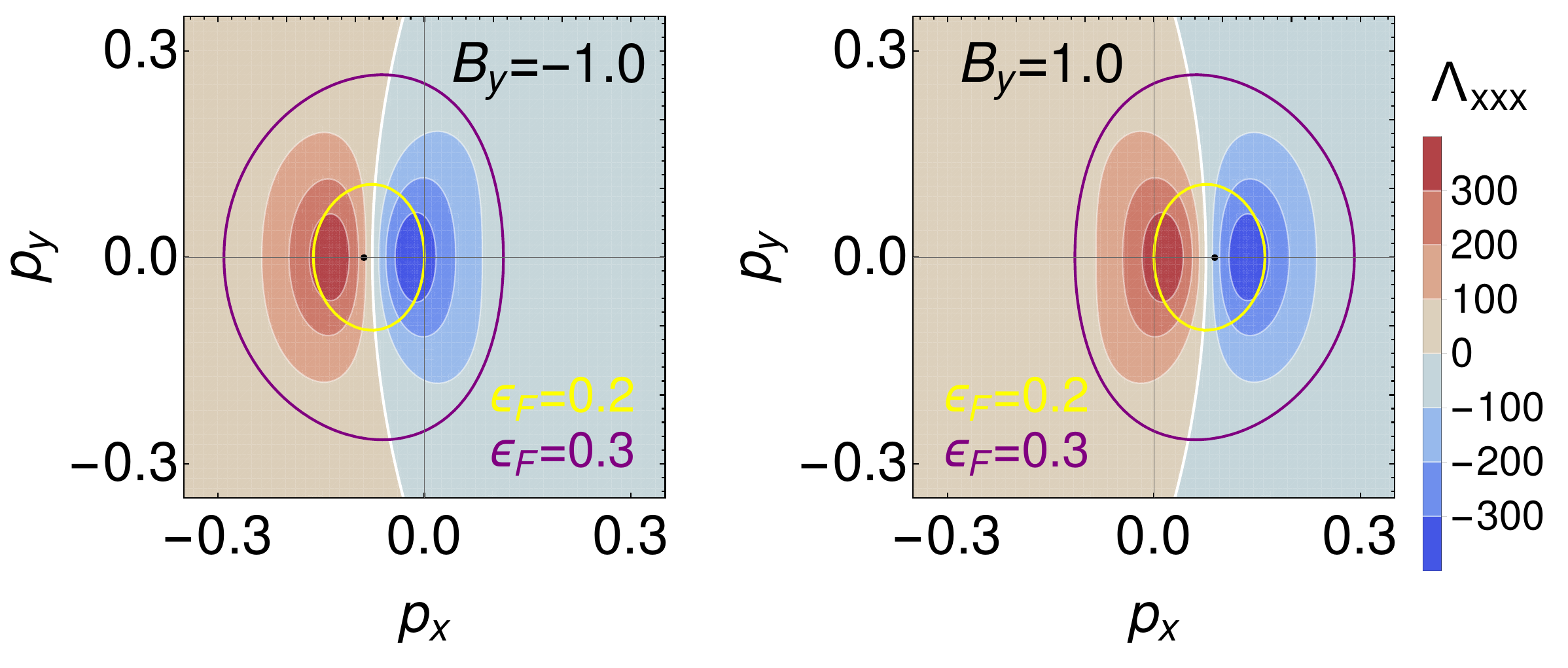}
    \caption{Contour plots of the band-energy normalized quantum metric dipole component $\Lambda_{xxx}$ in units of $1/(\hbar v_F k_0^4)$ with $k_0$ a reference momentum in the presence of a  planar magnetic field along the ${\hat y}$ direction of amplitude $B=-1.0$ (left) and $B=1.0$ (right) in units of $\hbar v_F k_0$. The trigonal warping parameter is set to $\lambda=0.2 \hbar v_F / k_0^2$. The yellow (purple) line marks the Fermi line for $\epsilon_F=0.2$ ($\epsilon_F=0.3$) measured in units of $\hbar v_F k_0$. The black dot marks the position of $p_x^0$. There is a mismatch between the center of the Fermi ellipses and the lines of zero (in white) of $\Lambda_{xxx}$.}
    \label{fig:figSM1}
    \end{center}
\end{figure}
Furthermore, and as shown in Fig.~\ref{fig:figSM1} the Fermi lines are ellipses centered on the mirror invariant line $p_y \equiv 0$ at the finite momentum 
$$p_x^0 \simeq \dfrac{3 B_y^5 \lambda^2}{\left(\hbar v_F\right)^7} \dfrac{1}{1 + 9 \dfrac{\lambda^2 B_y^4}{\left(\hbar v_F \right)^6}},$$
which is at larger and lower
momenta for positive and negative
values of the magnetic field.
%
This offset therefore leads to a net BNQM dipole which increases in magnitude with the Fermi energy until both high-intensity peaks are inside the Fermi line. This implies the existence of a Fermi energy sweet spot where the magnitude of the net BNQM dipole is maximal. 
The value of this optimal Fermi energy [see Fig.~\ref{fig:figSM3}(a1)] grows with the size of the externally applied planar magnetic field, 
differently from the maximal net BNQM dipole that is instead inversely proportional to the magnetic field amplitude. 
\begin{figure}[h!]
    \begin{center}
         (a)\includegraphics[width=0.55\textwidth]{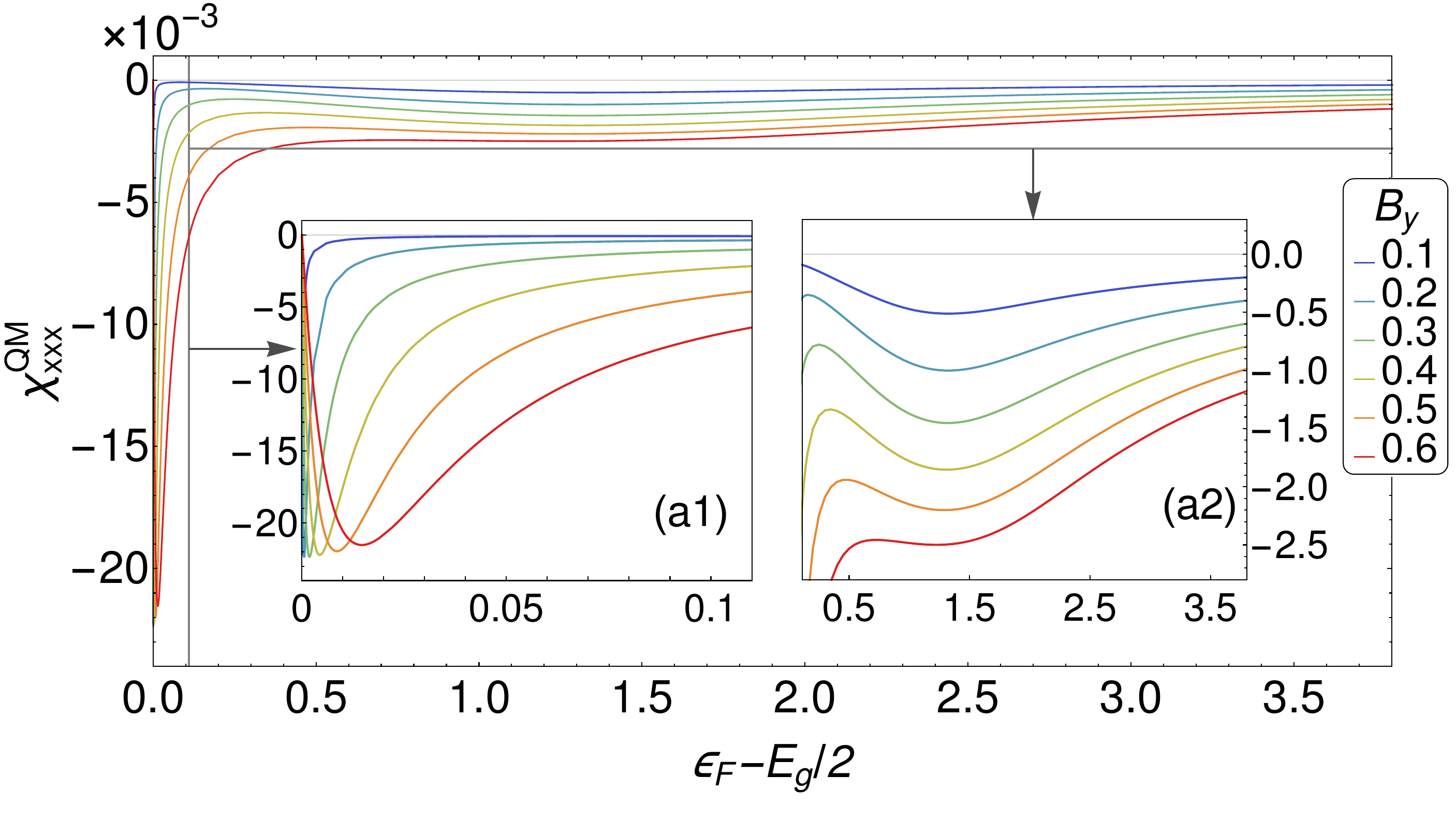} \hspace{0.15cm}
         \includegraphics[width=0.38\textwidth]{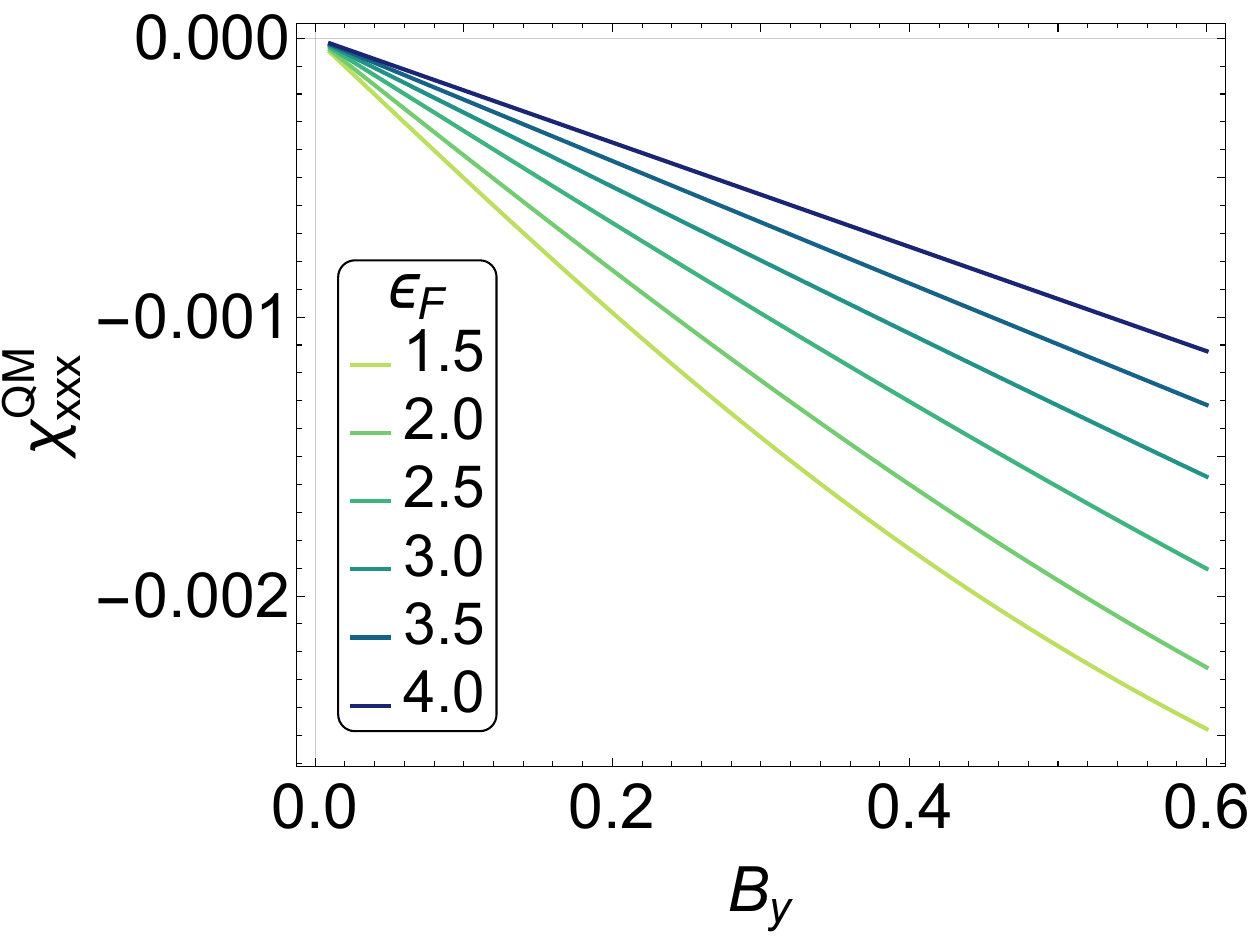}(b)
    \caption{(a) Behavior of the quantum metric induced nonlinear longitudinal conductivity $\chi_{xxx}^\text{QM}$ measured in units of $e^3 / (\hbar^2 v_F k_0^2)$ in a trigonal TI
  as a function of the Fermi energy $\epsilon_F$ measured in units of the reference energy $\hbar v_F k_0$ from the surface conduction band bottom  $E_g/2$, for different values of the planar magnetic field (in units of $\hbar v_F k_0$). The warping parameter is set to $\lambda=0.2~\hbar v_F / k_0^2$. The insets (a1)-(a2) show a zoom of  $\chi^\text{QM}_{xxx}$ in two different regimes. (b) Plot of QM-induced nonlinear conductivityin the regime of small applied field $B_y$ and intermediate densities keeping the Fermi energy constant.}
    \label{fig:figSM3}
    \end{center}
\end{figure}
At higher values of the Fermi energy the net BNQM dipole displays a downturn [see Fig.~\ref{fig:figSM3}(a2)] that signals a crossover to a different regime. In this region, the BNQM dipole magnitude increases almost linearly with the strength of the planar magnetic field [see Fig.~\ref{fig:figSM3}(b)]. This behavior can be understood by noticing that already at intermediate fillings the Fermi line can be approximated as a circle of radius $k_F$. In this case, the first non-vanishing contribution to the nonlinear response function can be found by expanding the density dipole component $\Lambda_{xxx}$ up to first order in $B$ and second order in the hexagonal warping strength $\lambda$. 

\begin{figure}[h!]
    \begin{center}
    \includegraphics[width=0.65\textwidth]{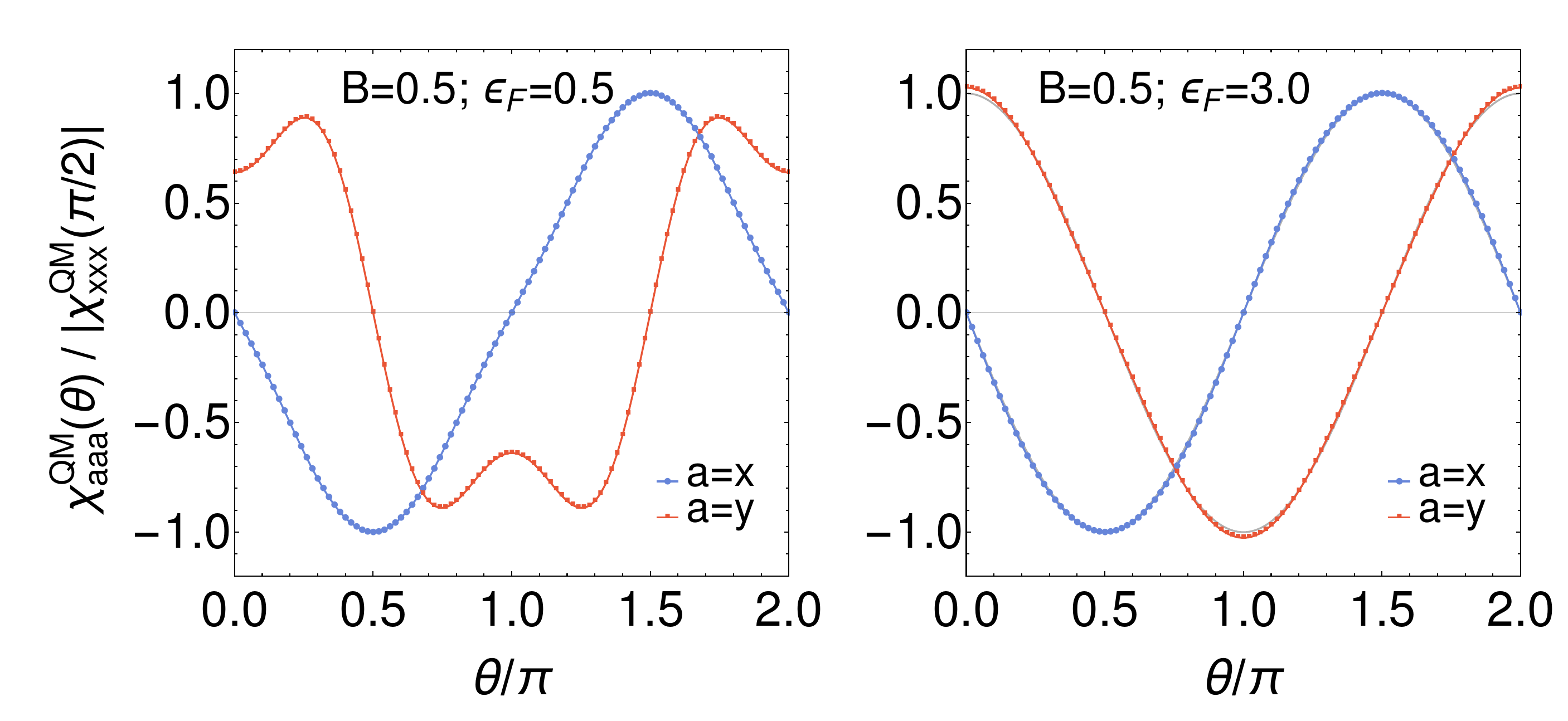} 
    \caption{Angular dependence of $\chi^\text{QM}_{xxx,yyy}$ by rotating the planar magnetic field for two different values of Fermi energy. Here $\theta$ is the angle of the magnetic field from the $\hat{x}$ principal crystallographic direction. In the right panel the gray lines indicate a simple sine (cosine) behavior.
    }
    \label{fig:figSM4}
    \end{center}
\end{figure}
Fig.~\ref{fig:figSM4} shows the behavior of the net BNQM dipole by rotating the magnetic field in the surface plane. As noted in the main text,  it exhibits a  $2\pi$ periodic, {\it i.e.} odd under a magnetic field reversal, angular dependence. 
We note that for a fixed direction of the magnetic field, topological surface states with right-handed ($v_F>0$) and left-handed ($v_F<0$) spin textures are characterized by opposite nonlinear longitudinal currents, and thus have an opposite angular dependence. 
Fig.~\ref{fig:figSM4} also displays the angular dependence of the net BNQM dipole governing the non-linear longitudinal conductivity $\chi_{yyy}^{\text{QM}}$, which is non-vanishing when the planar magnetic field has a finite component along the $\hat{x}$ direction. 
The related angular dependence is qualitatively different from $\chi_{xxx}^{\text{QM}}$ with a strongly pronounced anharmonic profile in the low-filling regime. 
This is due to the different electronic characteristic of the topological surface states when subject to a magnetic field in the $\hat{x}$ direction: a mirror symmetry preserving planar magnetic field does not gap the topological Dirac cone. As a result, even at low filling the non-linear longitudinal conductivity displays a linear increase with the Fermi energy 
At higher fillings $\chi_{xxx,yyy}^{\text{QM}}$ have both a purely harmonic profile [c.f. Fig.~\ref{fig:figSM4} right panel] and related to each other by a simple translation, similarly to the angular dependence of the semiclassical Drude contribution~\cite{he19}. 

\newpage
\section{Additional information on nonlinear magnetotransport in mercury chalcogenides grown along the $(001)$ direction}

\begin{figure}[h]
    \begin{center}
    (a)\includegraphics[width=0.4\textwidth]{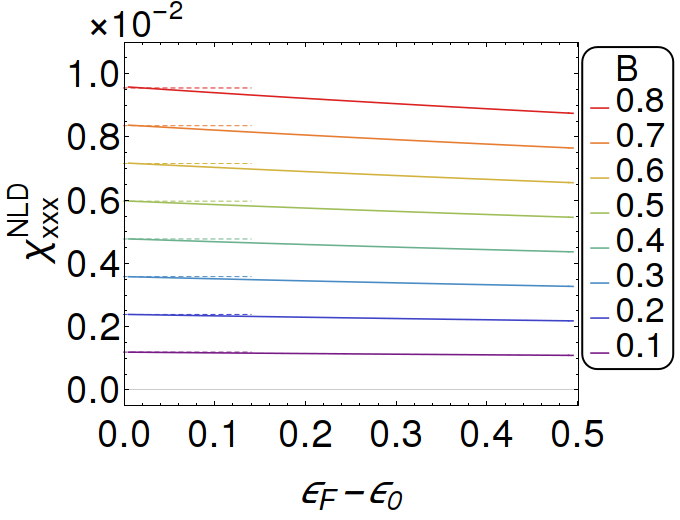}
         \hspace{0.55cm}(b)\includegraphics[width=0.4\textwidth]{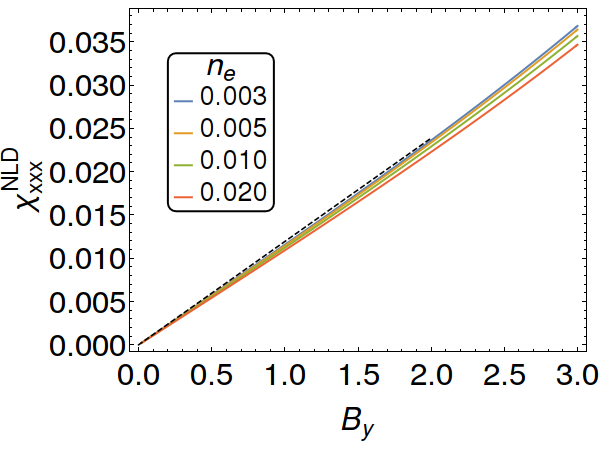}
    \caption{(a) Behavior of the nonlinear Drude conductivity $\chi_{xxx}^\text{NLD}$ measured in units of $e^3 \tau^2 v_F / \hbar^2$ as a function of the Fermi energy $\epsilon_F$ measured in units of $\hbar v_F k_0$ with $k_0$ a reference momentum at the $(001)$ surface of HgX compounds. We have subtracted the rigid shift $\epsilon_0= \alpha B^2 / (\hbar^2 v_F^2)$ from the Fermi energy. The dashed lines are the limiting values of Eq.~\eqref{eq:analytical}. (b) Behavior of the nonlinear Drude conductivity 
    as a function 
    of the amplitude of the applied planar magnetic field for constant surface carrier  density $n_e$ 
    The magnetic field amplitudes and energies are measured in units of $\hbar v_F k_0$ with $k_0$ a reference momentum. The density $n_e$ is measured in units of $k_0^2.$ The dashed black lines corresponds to Eq.\eqref{eq:analytical}.}
    \label{fig:HGTE}
    \end{center}
\end{figure}

In Fig.~\ref{fig:HGTE}(b-c) we show the behavior of the extrinsic Drude nonlinear conductivity in topological insulators with ${\mathcal C}_{2v}$ surface point group. The semiclassical term approaches a constant value when the Fermi energy approaches the Dirac point that is fixed at $\alpha B_y^2 / (\hbar^2 v_F^2)$. Following the procedure outlined in the main text we find 
\begin{equation}
\chi_{xxx}^\text{NLD} \simeq \dfrac{3 e^3 \tau^2 \alpha B_y}{8 \pi \hbar^4 v_F} 
\label{eq:analytical}
\end{equation}
We find a very good agreement between the expression above and the numerical results.

\begin{figure}[tb]
    \begin{center}
         (a) \includegraphics[width=0.38\textwidth]{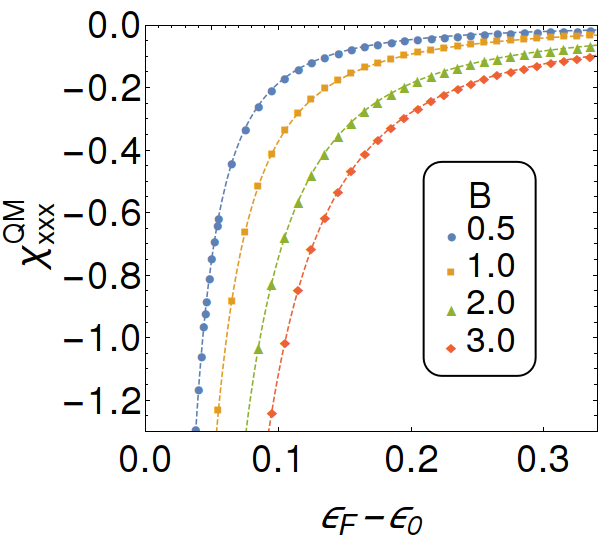}
         \hspace{0.54cm}(b)\includegraphics[width=0.46\textwidth]{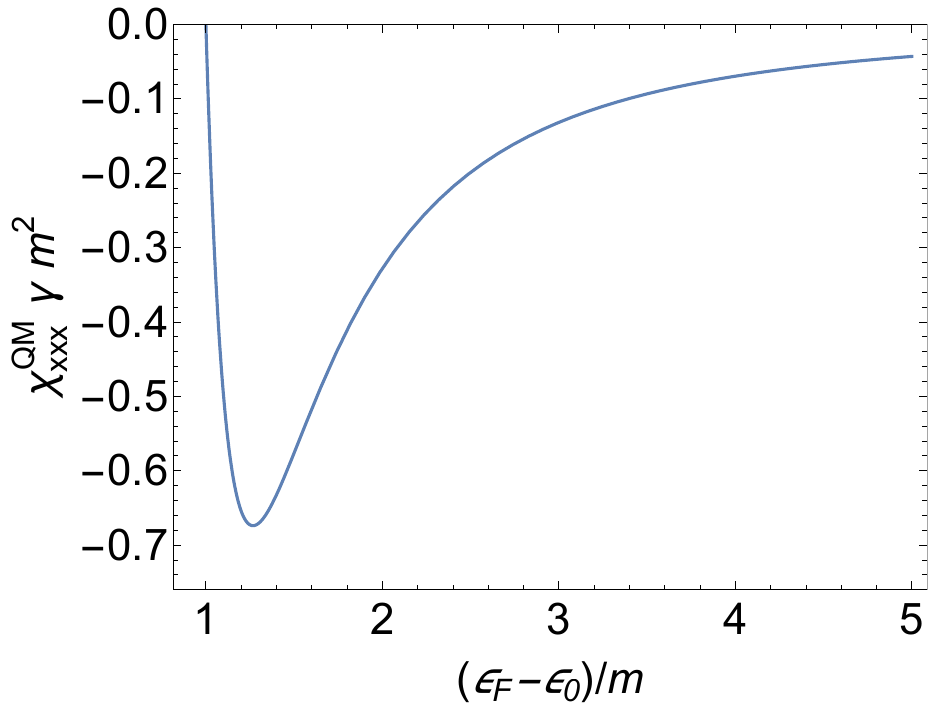}
    \caption{(a)Behavior of the QM-induced nonlinear conductivity measured in units of $e^3 / (\hbar^2 v_F k_0^2)$ as a function of the Fermi energy for different values of the planar magnetic fields (measured in units of $\hbar v_F k_0$). The Dirac point energy is $\epsilon_0 = \alpha B_y^2 / (\hbar^2 v_F^2)$.  The dashed lines are the analytical expression of Eq.~(3) of main text while the dots are numerical results. (b) 
    Same  for a massive Dirac cone as a function of the Fermi energy $\epsilon_F$ measured in units of the mass gap $m$ as obtained from Eq.~\ref{eq:qmapprox}. We have subtracted the Dirac point energy $\epsilon_0=\alpha B_y^2 / (\hbar^2 v_F^2)$. The conductivity $\chi_{xxx}^\text{QM}$ is measured in units of $1 / \gamma m^2$ where $\gamma= 128 \pi \hbar^2 v_F / (15 e^3 B_y \alpha)$.  }
    \label{fig:figSM7}
    \end{center}
\end{figure}

As shown Fig.~\ref{fig:figSM7}(a), the QM-induced nonlinear conductivity $\chi_{xxx}^{\text{QM}}$ displays instead a divergence as the Dirac point is approached.  We now show that such divergence is regularized once an infinitesimal mass due for instance to an out-of-plane component of the applied field or hybridization between the Dirac cones at the two opposite surfaces, is introduced. 
We thus consider the effective Hamiltonian
\begin{equation}
 {\mathcal H}_{\text{HgX}} = \hbar v_F \left(p_x \sigma_y  - p_y \sigma_x \right) + m \sigma_z - 2 \alpha \dfrac{p_x B_y}{\hbar v_F} \sigma_0+ \alpha \left(p_x^2 + p_y^2 \right) \sigma_0 + \dfrac{\alpha B_y^2}{(\hbar v_F)^2},
\end{equation}
where we have explicitly considered a magnetic field along the $\hat{y}$ direction. Using the same procedure outlined in the main text we find the analytical expression for the QM-induced nonlinear longitudinal conductivity 
\begin{equation}
\chi_{xxx}^{\text{QM}} \simeq  - \dfrac{15 e^3 B_y \alpha}{128 \pi \hbar^2 v_F} \left[ \dfrac{1}{ \left(\epsilon_F - \dfrac{\alpha B_y^2}{\hbar^2 v_F^2} \right)^2} + \dfrac{2 m^2}{ \left(\epsilon_F - \dfrac{\alpha B_y^2}{\hbar^2 v_F^2} \right)^4} - \dfrac{3 m^4}{ \left(\epsilon_F - \dfrac{\alpha B_y^2}{\hbar^2 v_F^2} \right)^6}\right]
\label{eq:qmapprox}
\end{equation}
Importantly and as shown in Fig.~\ref{fig:figSM7}(b) $\chi_{xxx} \rightarrow 0$ as the Fermi energy is at the bottom of the gapped Dirac bands, {\it i.e.} for $\epsilon_F - \alpha B_y^2 / (\hbar^2 v_F^2) \equiv m$. 

\begin{figure*}[h]
\centering
 \includegraphics[width=11cm,scale=1]{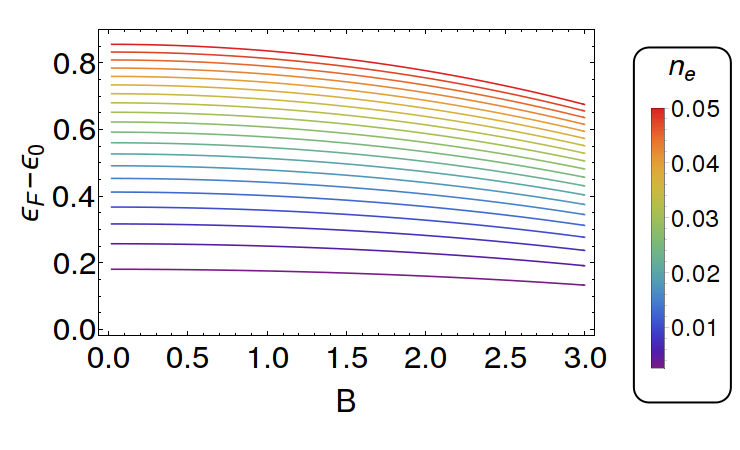}
	\caption{Fermi energy, measured from  the Dirac point $\epsilon_0= \alpha B^2 / (\hbar^2 v_F^2)$, as a function of the external magnetic field strength for different constant values of surface carrier density $n_e$. The magnetic field amplitudes and energies are measured in units of $\hbar v_F k_0$ with $k_0$ a reference momentum and  the density $n_e$ is measured in units of $k_0^2.$}
	\label{fig:carrierdensity_HGTE}
\end{figure*}

Fig.~\ref{fig:carrierdensity_HGTE} shows the magnetic field dependence of the Fermi energy, measured from the Dirac point energy, at fixed values of the surface carrier densities. In the weak magnetic field regime, the Fermi energy remains practically constant, guaranteeing the linear behavior of the intrinsic contribution to the nonlinear conductivity [c.f. Fig.4 of the main text]. At larger magnetic fields, we find a quadratic behavior of the Fermi energy that results in the nonlinearities of $\chi_{xxx}^{\text{QM}}.$

\newpage

\section{Nonlinear planar magnetotransport of trivial surface states}
\label{sec:rashba}
As mentioned in the main text, the generic feature of the QM-induced nonlinear conductivity from topological surface states is the absence of a sign change and hence a change from positive to negative nonlinear magnetoconductivity. 
We now show that topologically trivial surface states must display instead at least a sign change in the magnetoconductivity, which is consequence of the parity of the Fermi lines. 
To make things concrete, let us consider the surface states of strongly spin-orbit coupled materials, Bi for instance, which can be described by a general Rashba Hamiltonian 
\begin{equation}
\label{eq:HRashba}
    \mathcal{H}=\frac{\hbar^2 k^2}{2 m}+\alpha_R(k_x\sigma_y - k_y\sigma_x) + B_y\sigma_y
\end{equation}
\begin{figure*}[h]
\centering
 \includegraphics[width=15cm,scale=1]{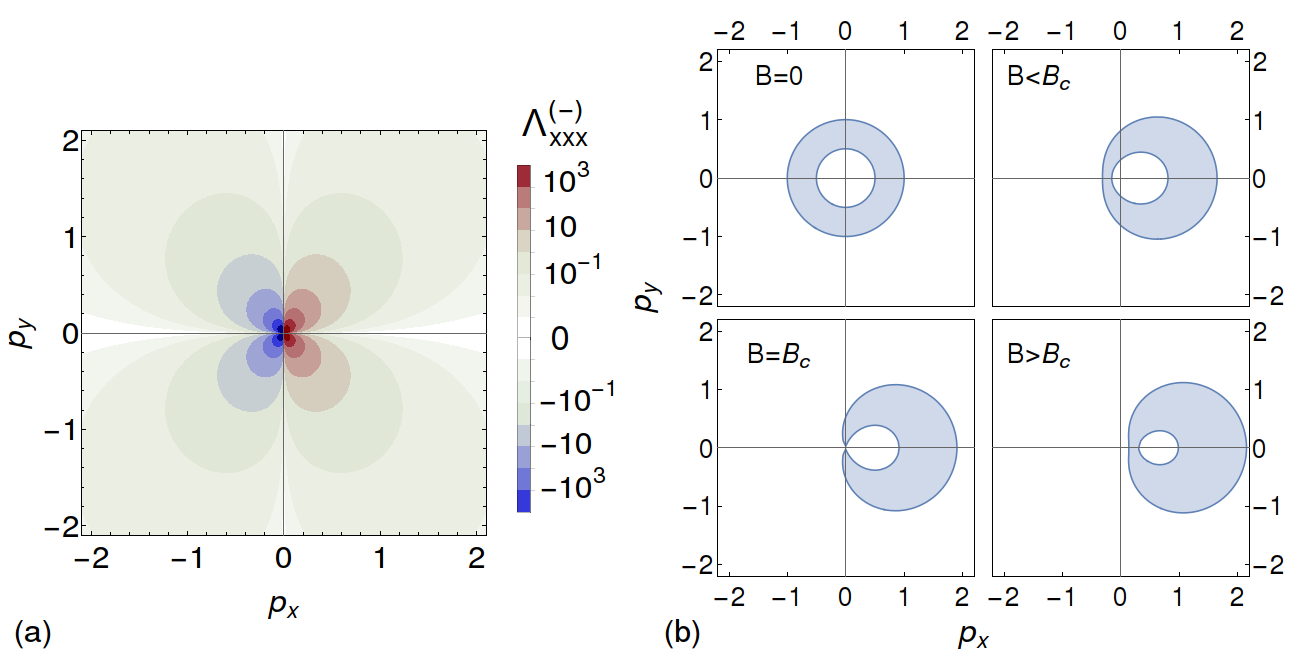}
	\caption{
    (a) Density plot of the band normalized quantum metric dipole $\Lambda_{xxx}$ of the outer spin-split Rashba band. 
    (b) Fermi lines varying the strength of the planar magnetic applied along the $\hat{y}$ direction for constant electron density, specifically in the plots it is $n_e=0.2 k_0^2 $, where   $k_0$ is a reference momentum.  The shaded region between the two Fermi lines is the integration region to evaluate $\chi_{xxx}^\text{QM}$. The Rashba parameter here is set to $\alpha_R=0.5 k_0/E_0$, with $E_0=\hbar^2 k_0^2/2m$, the critical field is $B_c
    \simeq 0.355 E_0$. The other two values of the field $B$ used in the plot are $B=0.25 E_0$ and $B=0.45 E_0$.
    }
	\label{fig:Fermi_annulus}
\end{figure*}
At each value of ${\bf k}$ the BNQM dipole density of the two spin-split bands cancel each other. However, there is a region of crystal momenta populated by a single spin band. This region, which corresponds to the annulus between the two Fermi lines of the system, has a nonvanishing BNQM dipole density  $\Lambda_{xxx}$ that gives a rise to a net BNQM dipole and consequently an intrinsic contribution to the nonlinear conductivity. 
To derive its main features, we introduce as for the case of single Dirac cones at the TI surfaces the shifted momenta $p_x=k_x + B_y / \alpha_R$ and $p_y=k_y$. In terms of these, the BNQM dipole density is constant in $B_y$ \cite{sal24}: all the magnetic field dependence is in the shape of the Fermi lines for the outer and inner branch of the Rashba bands. 
As shown in Fig.~\ref{fig:Fermi_annulus}(a) $\Lambda_{xxx}$ is odd in $p_x$ which implies together with the circular shape of the uncompensated annulus [c.f. Fig.~\ref{fig:Fermi_annulus}(b)] the absence of a finite $\chi_{xxx}^{\text{QM}}$ at $B=0$. 
At weak magnetic field values, the uncompensated annulus is shifted at larger values of $p_x$ [c.f. Fig.~\ref{fig:Fermi_annulus}(b)].
\begin{figure*}[h]
\centering
(a) \includegraphics[width=0.4\textwidth]{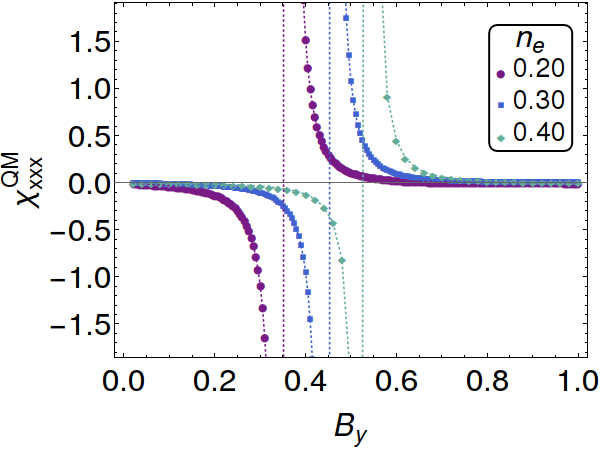}\hspace{0.3cm}
(b)  \includegraphics[width=0.4\textwidth]{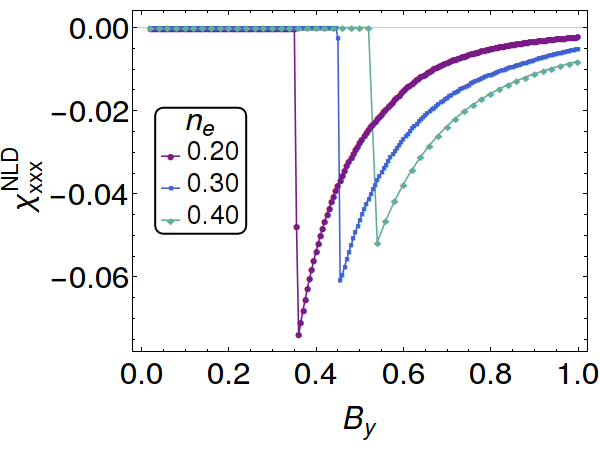}
	\caption{Behavior of the quantum metric-induced (a) and non-linear Drude (b) longitudinal  conductivity of Rashba surface states 
    for three different values of the electronic density. The Rashba parameter used in the plots is $\alpha_R=0.5$.} 
	\label{fig:Rashba_chixxx}
\end{figure*}
The net BNQM dipole then assumes negative values [c.f. Fig.~\ref{fig:Rashba_chixxx}(a)] since shifted momenta with $p_x<0$ closer to the origin lie in the uncompensated annulus -- the BNQM dipole density is diverging at the shifted momenta origin. 
At  a critical magnetic field $B_c$,  the uncompensated annulus touches this singular point and $\chi_{xxx}^{\text{QM}}$ diverges. Furthermore, the divergence is negative or positive depending if the magnetic field approach $B_c$ from below or above respectively [c.f. Fig.~\ref{fig:Rashba_chixxx}(a)]. 
Importantly, this divergence only occurs at zero temperature. Finite temperature effects will lead to an overall smearing with the nonlinear conductivity profile that will display a double peak with a sign change at the critical magnetic field.
Finally, in the strong magnetic field regime the net BNQM dipole is positive as the uncompensated annulus assume only $p_x>0$ values [c.f. Fig.~\ref{fig:Fermi_annulus}(b)]. Therefore, for weak and strong magnetic fields, $\chi_{xxx}^{\text{QM}}$ assumes values of opposite signs which requires the presence of an odd number of sign changes. 

